\documentstyle[eqsecnum,aps]{revtex}

\begin{document}
\draft
\preprint{IHEP 98-67}
\title{On the value of $\alpha_s$ from 
the analysis of the SLAC/BCDMS deep inelastic scattering data.}
\author{S. I. Alekhin}
\address{Institute for High Energy Physics, Protvino, 142284, Russia}
\date{October 1998}
\maketitle
\begin{abstract}  
We performed NLO QCD analysis of the nonsinglet part of 
the combined SLAC/BCDMS 
data on $F_2$ with the extraction of $\alpha_s$ and high twist 
contribution. It was shown that the value of $\alpha_s$ obtained 
in the analysis is sensitive to the statistical inference 
procedures dealing with systematic errors on the data.
The fit with the complete account of point-to-point correlations 
of the data gave the value of $\alpha_s(M_Z)=0.1180\pm0.0017 (68\% C.L.)$,
to be compared with the previously reported value 
of $\alpha_s(M_Z)=0.113\pm0.003 (99\% C.L.)$.
This new value of $\alpha_s$ is 
compatible with the  LEP measurements and the world average. 
The high twist contribution being
strongly anti-correlated with the value of $\alpha_s$
became lower than it was previously reported.
\end{abstract}

\section{Introduction}

It is well known that the value of the strong coupling constant
$\alpha_s(M_Z)$ measured at LEP  is larger than the
value 
of $\alpha_s(M_Z)$ obtained from the evolution of  
results of the analysis of the combined SLAC/BCDMS 
DIS data on proton and deuterium 
\cite{VIR}
laying at lower $Q^2$ \cite{LEP}. 
This discrepancy  
caused a lot of discussions (see e.g. \cite{SHIF})
and is often attributed to the existence of 
new fundamental 
particles, which can change the dependence of $\alpha_s$ on $Q^2$. 
Meanwhile, the value of $\alpha_s$ from \cite{VIR}
is strongly correlated with the value of 
simultaneously fitted high twist (HT)
contribution.
This correlation is inevitable if one does not make
a sufficient $Q^2$ cut of data, otherwise the 
power corrections can essentially, if not completely,
imitate the logarithmic scaling violation \cite{POWER}.
The separation of the power and logarithmic behavior
is complicated in the case of SLAC/BCDMS data analysis
because these data do not practically overlap and
exhibit significant discrepancies in the vicinity of the overlapping
regions. To achieve a satisfactory description of the data,
one is to invent a method to interpret these discrepancies,
which is obviously cannot be done without some adoptions. 
The larger is the correlation of the fitted  parameters
the more sensitive their values are to the perturbations 
of other inputs to the fit and hence any adoption 
made in the analysis should be accurately clarified.
The analysis \cite{VIR} is not absolutely rigorous 
in the points concerning 
the inference of systematic errors.
The number of independent systematic errors for the combined SLAC/BCDMS
data set is about 40 and the authors of \cite{VIR} 
combined most of them in a quadrature claiming that this
would not distort the results.
In the present work we investigated the effect of this adoption
on the bias of the fitted parameters. 

\section{The data and their systematic errors}
We analysed essentially the same data set
\cite{WHIT,BCDMS} as in \cite{VIR}
with the minor differences:
\begin{itemize}
\item
  we used the data on cross sections separated by the beam energies
  instead of merged data on $F_2$. For the SLAC data we withdraw 
  the merging errors in this way. The BCDMS data within
  this approach were reduced 
  to the value of $R=\sigma_L/\sigma_T$ \cite{WHIT} common 
  to the SLAC data.
\item
  we imposed the more stringent cut $x\ge0.3$
  to prevent additional uncertainties due to a poorly 
  known gluon distribution. This cut leaves 
  the data which can be in good approximation described by 
  the pure nonsinglet structure functions, which essentially
  reduces the number of the fitted parameters.
  At the same time the value of $\alpha_s$ in the fit to
  the combined SLAC/BCDMS data is basically determined by 
  the high-x points and we did not loose statistical 
  significance of the analysis as one can see from the final 
  results.
  The cut $x\le0.75$ coinciding with \cite{VIR} 
  and rejecting the region where
  the binding effects in deuterium can be important
  was also imposed in the analysis. The $Q^2$ range of the data
  left after the cut is $1~GeV^2<Q^2<230~GeV^2$.
\end{itemize}
The number of data points (NDP) and the number of independent systematic 
errors (NSE)
for each experiment used in our analysis
are presented in Table I. The systematic errors on the BCDMS data 
are presented by the following
independent sources: calibration of the measurement of 
the incident and scattered muon energy, resolution of the spectrometer,
detector and trigger inefficiencies, relative normalization of 
data from internal and external targets, general normalization
and relative normalization uncertainties between 
the data set taken at different beam energies. 
(The latest were ascribed to
the data at beam energies of 
100, 120 and 280 GeV while the data at 200 GeV
were considered as the reference ones.)
In the analysis \cite{VIR} the systematic errors
from the first three sources were combined in quadrature
into a single error called a ``main systematic error'' and
the data points were shifted by the value proportional 
to this combination
while the proportionality coefficient
was determined from $\chi^2$
minimization. The general normalization was also
considered as a free parameter and then the value of normalization
uncertainty presented in the source paper \cite{BCDMS}
was not explicitly accounted for. The rest systematic 
errors were considered as uncorrelated and were combined in quadrature with 
statistical errors. 

The correlated systematic errors on the  SLAC 
data arose due to: background contamination, spectrometer acceptance
uncertainties and radiative corrections uncertainties. In addition, 
as far the older SLAC data were normalized to the data from 
the E-140 experiment, there are two more systematic errors on them:
target dependent and target independent relative 
normalization uncertainties. (The data from E-140 experiment have 
only one additional absolute normalization error). In the analysis
\cite {VIR} all these errors 
were combined in quadrature with statistical 
ones.

\section{Fitted formula}
The QCD input leading twist (LT)
structure functions of 
proton and neutron were
parametrized at the starting 
value of $Q_0^2=9 GeV^2$ as follows:
\footnote{We checked that extra polynomial-type
 factors do not improve 
the quality of the fits.}
\begin{displaymath}  
F^p_2(x,Q_0)=A_{p}x^{a_{p}}(1-x)^{b_{p}}\frac{2}{N_p}
\end{displaymath}  
\begin{displaymath}  
F^n_2(x,Q_0)=A_{n}x^{a_{n}}(1-x)^{b_{n}}\frac{1}{N_n}.
\end{displaymath}  
Here conventional normalization factors $N_p$ and $N_n$ are
\begin{displaymath}  
N_{p,n}=\int_0^1dxx^{a_{p,n}-1}(1-x)^{b_{p,n}}.
\end{displaymath}  
These distributions were evolved through the region of $Q^2$ occupied 
by the data in NLO QCD approximation 
within $\overline{MS}$ factorization scheme \cite{MS}
with the help of the code used earlier \cite{PDF96}. 
The $Q^2$ dependence of $\alpha_s$
was calculated as the numerical solution
of the equation 
\begin{equation}
\frac{1}{\alpha_s(Q)}-\frac{1}{\alpha_s(M_Z)}=
\frac{\beta_0}{2\pi}\ln\biggl(\frac{Q}{M_Z}\biggr)+
\beta\ln\biggl[\frac{\beta+1/\alpha_s(Q)}{\beta+1/\alpha_s(M_Z)}\biggr],
\label{ALPHA}
\end{equation}
where
\begin{displaymath}
\beta_0=11-\frac{2}{3}n_f,~~~~\beta=\frac{2\pi\beta_0}{51-\frac{19}{3}n_f}
\end{displaymath}
and 
the number of the active fermions $n_f$ was changed 
at the values of
$Q$ equal to quark masses keeping the continuity of $\alpha_s$.
The final formula for structure function used in the fit with account 
of twist-4 contribution
was choose the same as in \cite{VIR}:
\begin{displaymath}  
F_2^{(P,D),HT}=F_2^{(P,D),LT}\Bigl[1+\frac{h^{(P,D)}(x)}{Q^2}\Bigr],
\end{displaymath}  
where $F_2^{(P,D),LT}$ are the leading twist terms
with account of the target mass correction \cite{TMC}.
The functions $h^{(P,D)}(x)$ were parametrized in the model independent 
way: their values at $x=0.3,0.4,0,5,0.6,0.7,0.8$ were fitted,
between these points the functions were linearly interpolated.
As we mentioned before,
we used the common value of $R$ \cite{WHIT}
for all the data including BCDMS ones.

\section{Results}
\subsection{BCDMS reanalysis}
At the first stage of our analysis we used the inference procedures
analogous to \cite{VIR}. The parameters were evaluated through 
minimization of the functional 
\begin{equation}  
\chi^2=\sum_{K,i}
\frac{\Bigl[(f_i-\lambda_K\Delta y_i)/\xi_K-y_i\Bigr]^2}{\sigma_i^2},
\label{SIMPL}
\end{equation}  
where $K$ runs through the 
data subsets obtained by separation of all analysed data on 
experiments and 
targets; $i$ - through data points
within these subsets.
The other notations are:
$y_i$ - the measurements, $\sigma_i$ - the statistical errors, combined
with some systematic errors as described above, $f_i$ - theoretical model 
prediction depending on the fitted parameters, $\Delta y_i$ -
the ``main systematic
error'' on the BCDMS data, $\lambda_K$ and $\xi_K$ are fixed at 0. and 1., 
correspondingly for the
SLAC experiments and are the fitted parameters for BCDMS.
For the test purposes we fitted formula with the parameters $\xi$
and $\lambda$ fixed at their values as given in \cite{VIR}.
The obtained results are presented in column 1 of Table II and on Fig. 1.
The values of HT coefficients obtained in the analysis \cite{VIR}
are also presented on Fig. 1. As far the errors quoted for them
in \cite{VIR} correspond to the change of $\chi^2$ equal to 9., their
pictured errors are scaled by the factor of 1/3 to provide a meaningful
comparison with our figures.
One can see that they coincide within the statistical fluctuations.

The next step was to release these parameters (the results 
are presented in column 2 of Table II).
We can note that for this fit the BCDMS 
data are renormalized slightly smaller. As a consequence, the value of
$\alpha_s$, which exhibits negative correlation with this normalization
factor became slightly less than that in \cite{VIR}.
In this connection note that one could suppose 
the dependence of the normalization 
factor on the x-cut 
because the x-shape of 
the BCDMS data does not 
match the SLAC data very well 
(in particular, it was pointed in \cite {WHITLTH}).
The errors of the $\alpha_s$ value increased two times
comparing with the first fit. This is in accordance with 
the above observation, that $\alpha_s$ is strongly correlated with the 
normalization factors for the BCDMS data - releasing the latest 
we allowed more room for the $\alpha_s$ variation.

An alternative possibility to account for the normalization 
error of the data is to introduce the correlation matrix
\begin{displaymath}
C_{ij}=\delta_{ij}\sigma_i\sigma_j+f_if_js_K^2
\end{displaymath}
into the minimized functional in the following way:
\begin{equation}
\chi^2=\sum_{K,i,j}
\Bigl[(f_i-\lambda_K\Delta y_i)-y_i\Bigr]E_{ij}
\Bigl[(f_j-\lambda_K\Delta y_j)-y_j\Bigr],
\label{NORM}
\end{equation}
where $s_K$ is the data normalization uncertainty 
for each target as it is estimated 
by the experimentalists and $E_{ij}$ is the inverse of $C_{ij}$;
$j$ runs through the data points of each data subset, 
$\delta_{ij}$ is the Kronecker symbol and the other notations
are the same as in (\ref{SIMPL}). This approach is natural if 
one considers a systematic error as a random variable, i.e.
within the Bayesian approach (see more in \cite{BAY} on this scope). 
The fit within this approach is, in principle,
more stable comparing with the 
renormalization approach (\ref{SIMPL}) because in (\ref{NORM}) 
the normalization parameter variation is limited by the
scale of $s$.
In our particular
case this is not so important as far 
one can see from Table II, that the normalization factors 
for the BCDMS data are anyway within their
normalization systematic error (3\%).
This anticipation is supported by the results of the fit 
within the approach (\ref{NORM}) which are
also presented in Table II (column 3).
Analogously the fitted parameters should not be sensitive to the 
the stabilization term $(\xi-1)^2/s^2$ added to functional (\ref{SIMPL})
in \cite{VIR}
as far this term corresponds effectively to the additional measurement
of $\xi$ with the average of 1. and the error of $s$; 
the weighted averaging of this 
measurement with the value of $\xi$ from Table II
cannot evidently change the latest one.

To proceed with the implementation of Bayesian approach 
for the treatment of systematic errors, we minimized 
the functional 
\begin{equation}
\chi^2=\sum_{K,i,j}
(f_i-y_i)E_{ij}(f_j-y_j),
\label{MAIN}
\end{equation}
where $E_{ij}$ is the inverse of the correlation matrix
\begin{displaymath}
C_{ij}=\delta_{ij}\sigma_i\sigma_j+f_if_j(\vec{s}_i^K \cdot \vec{s}_j^K)
\end{displaymath}
and each 4-component
vector $\vec s_i^K$ includes the normalization uncertainty
as well as the three systematic errors which were initially combined into 
the ``main systematic error''
of the BCDMS data.
The most interesting difference of 
this fit results (presented in 
Table II, column 4) from the previous fits is the increase of $\alpha_s$.
The value of $\alpha_s$ is strongly anticorrelated 
with the HT contribution 
at high $x$ and naturally the last-named
decreases correspondingly.
The effect is of the order of one standard deviation (as could be  
anticipated because the value of $\lambda$ is of the order of 1.
when it is released in the fit), with the tendency to decrease the 
discrepancy with the LEP data. Alongside one can observe the 
decrease of $\chi^2$, which is connected with the fact that 
in the earlier fits main systematic errors were, as a whole, underestimated 
when combined in quadrature.

An additional improvement is to account, within this approach, for 
two more BCDMS
systematic errors, which were not included in the ``main systematics'':
The errors due to detector and trigger inefficiencies.  
The results of this fit are presented in column 5 of 
Table II. 
Again we can see the 
enlargement of $\alpha_s$ value and the correlated decrease of the HT
contribution, although not so large as in the case of 
the re-account of ``main systematics''.

The next step of our analysis was to  re-account the 
errors corresponding to the uncertainty in the relative 
normalizations of the data subsets 
for different energies. The results are presented 
in column 6 of Table II. The value of $\alpha_S$ again increased 
and the effect is even more pronounced than in the case 
with the re-account of ``main systematics''. This is not surprising
because as was stated by the BCDMS collaboration itself 
the uncertainty in the relative normalizations 
have the most effect on the error of $\alpha_s$ \cite{BCDMSQCD}.

Our final exercise with the BCDMS data 
concerns the correlation of systematic errors on the data from  
the proton and deuterium targets. 
The authors note that this 
correlation is large, but do  
not quantify it. To investigate the scale of this correlation effect,
we performed one more fit assuming the total correlation
(column 7 of Table II). The parameter estimates for real 
proton/deuterium correlation lie between the values from 
column 6 and 7, more close to 7 and we again observe
the increase of $\alpha_s$. 

Summarazing, we can conclude 
that a complete account of point-to-point 
correlations due to systematic errors on the BCDMS data
in the combined SLAC/BCDMS analysis
cancels the discrepancy with the LEP results.
The effect of $\alpha_s$ increase 
comparing with the previous analysis \cite{VIR}
arises mainly due to  
re-account of ``main systematics'' and the errors 
due to relative normalizations of 
the data taken at different energies.

\subsection{SLAC reanalysis}
For the completeness we 
accounted for the point-to-point correlation of 
the SLAC data too. At first we proceeded with 
the systematic errors on the
E-140 data only. The results of the fit are 
presented in column 1 of Table III and do not  essentially differ
from the previous fit. 
As mentioned above the older SLAC data were renormalized 
to the data from E-140 experiment  \cite{WHITLTH}.
Due to the absence of E-140 proton data 
the renormalization of proton data subsets  
was performed using ``bridging'' through the E-49B experiment,
which introduced additional uncertainties. As far we used 
more of the proton
data in the analysis, we preferred to perform the 
independent renormalization. 
Then, we removed from the systematic errors
on the older SLAC data
the relative normalization 
uncertainties which arose
due to their renormalization 
to E-140 and introduced the fitted normalization parameters
for each experiment and target into the functional (\ref{MAIN}): 
\begin{displaymath}
\chi^2=\sum_{K,i,j}
(f_i/\xi_K-y_i)E_{ij}(f_j/\xi_K-y_j), 
\end{displaymath}
where $\xi_K$ are fixed at 1. for the BCDMS and E-140 data subsets.
The results of this fit are presented in Table III, column 2.
One can see that our renormalization factors are,
as a whole, compatible 
with 1. within the errors, although there is some tendency to 
shift proton data up comparing with \cite{WHITLTH}.

The final step of our analysis was the incorporation of the rest 
systematic errors into the correlation matrix. The results of
this fit are presented in column 3, Table III. 
The value of $\alpha_s$ due to the 
last improvement remained unchanged, 
the main effect was a certain increase of $\chi^2$, 
while the statistical confidence of the fit remains good.
This is readily understood 
because if one combines the correlated errors in 
quadratures, the $\chi^2$ is underestimated.
In the final fit the relative normalization of SLAC data is
in the range of few percent up comparing with the BCDMS data.
In the global fits  
the SLAC data are often used as the reference
ones and the BCDMS data are renormalized to them and usually
are shifted down by few percent. Our renormalization scheme
is in principle compatible with the commonly used one, except for
the general normalization. This discrepancy cannot be clarified 
if one uses in the analysis only the data on DIS
as far it is well known that they cannot define the absolute
normalization parameters
very well, moreover, we applied the cut on $x$ in the analysis.
 Anyway, it is obvious, that the ambiguity in  the 
general absolute normalization cannot affect 
determination of a slope on $Q^2$ and, hence, change the 
value of $\alpha_s$.

\section{Summary}

The final value of $\alpha_s(M_Z)$ obtained in our analysis 
is presented in column 3 of Table III: 
\begin{displaymath}  
\alpha_s(M_Z)=0.1180\pm0.0017(stat+syst).
\end{displaymath}  
It is compatible with the values obtained in the LEP 
experiments \cite {LEP} and in the analysis of 
CCFR data on $F_3$ \cite{CCFR}
with the extraction of HT contribution
\cite{KAT}, but is in certain contradiction with the 
results of \cite{VIR}:
\begin{displaymath}  
\alpha_s(M_Z)=0.113\pm0.003(stat+syst).
\end{displaymath}  
For a meaningful comparison it is worth to remind that 
in the last result the error corresponds to 
the change of $\chi^2=9$, i.e. 
three standard deviations and, consequently, the 
distance between our result and \cite{VIR} is about
3 standard deviations.
The statistical confidence of our final
fit ($\chi^2/NDP=1179/1183$) is perfect,
while in \cite{VIR} $\chi^2/DOF=599/687$
and hence 
the value of $\chi^2$
is by more than two standard deviations lower 
than its supposed mean. This is yet within possible statistical
fluctuations, but nevertheless can signal about underestimation
of $\chi^2$ due to the combining of systematic errors in quadrature.

In \cite{VIR} the value of $\alpha_s(50~GeV)=0.180\pm0.008$
with the help of the approximate solution of (\ref{ALPHA})
\begin{displaymath}  
\alpha_s(Q)=\frac{2\pi}{\beta_0\ln(Q/\Lambda)}
\Bigl[1-\frac{2\pi}{\beta_0\beta}
\frac{\ln(2\ln(Q/\Lambda)}
{\ln(Q/\Lambda)}\Bigr] 
\end{displaymath}  
was transformed into the value of
\begin{displaymath}  
\Lambda^{(4)}_{\overline{MS}}=263\pm42(stat+syst)~MeV.
\end{displaymath}  
Our value of $\alpha_s(50 GeV)=0.1935\pm0.0048$
can be analogously transformed into
\begin{displaymath}  
\Lambda^{(4)}_{\overline{MS}}=337\pm28(stat+syst)~MeV.
\end{displaymath}  

The correlation matrix of the fitted parameters is 
presented in Table IV.
\footnote{We omitted the correlation coefficients corresponding to 
the normalization parameters of the SLAC data due to  
space limitation. The full correlation 
matrix can be obtained from the author on the request.} 
One can see from the table that the correlation 
of $\alpha_s$ with the HT coefficients 
is rather large. This supports our initial statement that the 
separation of logarithmic and power effects in a scaling violation is 
unstable under various assumptions.
Other effects, not taken into account before (e.g.
nuclear effects in deuterium), should be investigated before 
one can elaborate 
reliable estimate of $\alpha_s$ from 
the analysis of these data. 

As far the HT contribution and the value 
of $\alpha_s$ are
strongly anticorrelated, the increase of $\alpha_s$, which we 
observed above, is accompanied by the decrease of HT
\footnote{This effect was also recently observed in the analysis
\cite{BODEK}, where $\alpha_s(M_Z)$ was fixed at the value
of 0.120}.
The total effect on the HT magnitude is about a factor of 3/4, 
comparing with the results \cite{VIR}. In this connection it 
is interesting to compare our results with the predictions of 
the infrared 
renormalon (IRR) model \cite{WEBBER,STEIN}. This model is known to 
reasonably 
reproduce the shape of HT contribution obtained in 
\cite{VIR}, but the 
absolute value prediction is about 2.5 times higher 
than the data (see \cite{STEIN}). The comparison of our results 
with the IRR model predictions is presented on Fig. 2.
The model calculations were made in 
the nonsinglet approximation using the structure functions 
and the value of $\Lambda^{(4)}_{\overline{MS}}$
obtained in our analysis:
\begin{displaymath}  
h(x)=\frac{A'_2}{F_2^{LT}(x,Q)}\int_x^1dz
C_2(z)F_2^{LT}(x/z,Q)
\end{displaymath}  
\begin{displaymath}  
C_2(z)=-\frac{4}{(1-z)_+}+2(2+z+6z^2)-9\delta(1-z)-\delta'(1-z)
\end{displaymath}  
\begin{displaymath}  
A'_2=-\frac{2C_F}{\beta_0}\bigl[\Lambda^{(4)}_{\overline{MS}}\bigr]^2e^{-C},
\end{displaymath}  
where $Q^2=9~GeV^2$, $C_F=4/3$, $C=-5/3$ and $F_2^{LT}$ does not include 
target mass corrections here.
One can see
the improved agreement of the IRR model 
predictions with the data at $x=0.5-0.7$.

In conclusion, the separation of the logarithmic and power 
scaling violation effects in the analysis of deep inelastic 
scattering data is unstable due to a high correlation 
of these effects in the $Q^2$ region where they are 
both not small. The complete account of 
point-to-point correlations of the data lead to the 
shift in the value of $\alpha_s$ by about 3 standard 
deviations comparing with the simplified statistical 
inference procedure.
The HT contribution, which is strongly anti-correlated with $\alpha_s$,
decreases within this approach
and becomes more compatible with the 
prediction of IRR model at moderate x.
Further investigation of a possible perturbation in the 
analysis of DIS data is needed before a reliable value of $\alpha_s$
can be determined.

\acknowledgments
The author is indebted to A.L.Kataev and A.V.Sidorov for 
interesting discussions and E.Stein for reading the manuscript. 
The work was supported by RFFI grant 96-02-18897.

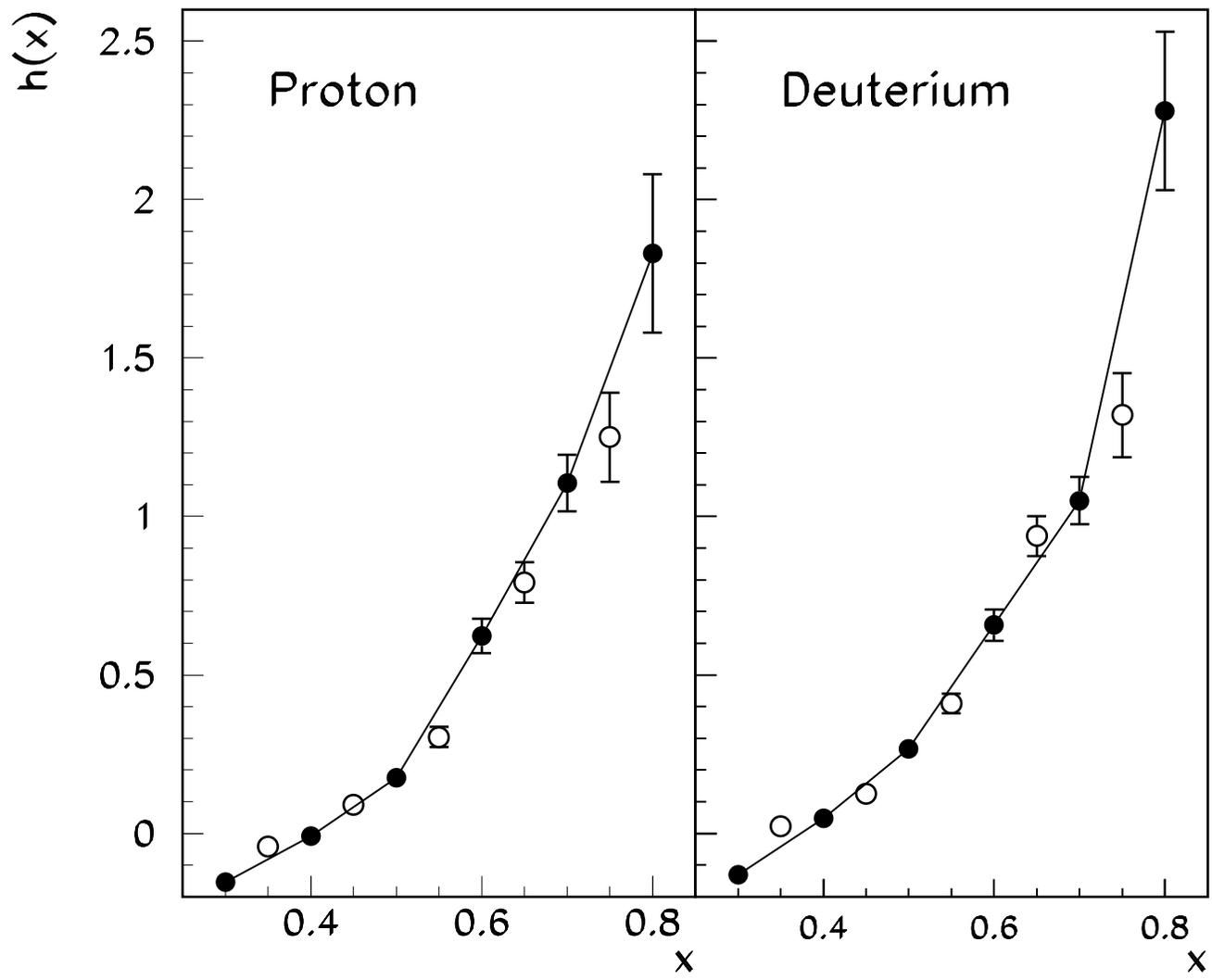
\begin{figure}
\caption{The high-twist contributions obtained in the fit 
with the functional (1) (full circles and lines).
The results of the analysis [2]
are presented for comparison (open circles).}
\end{figure}

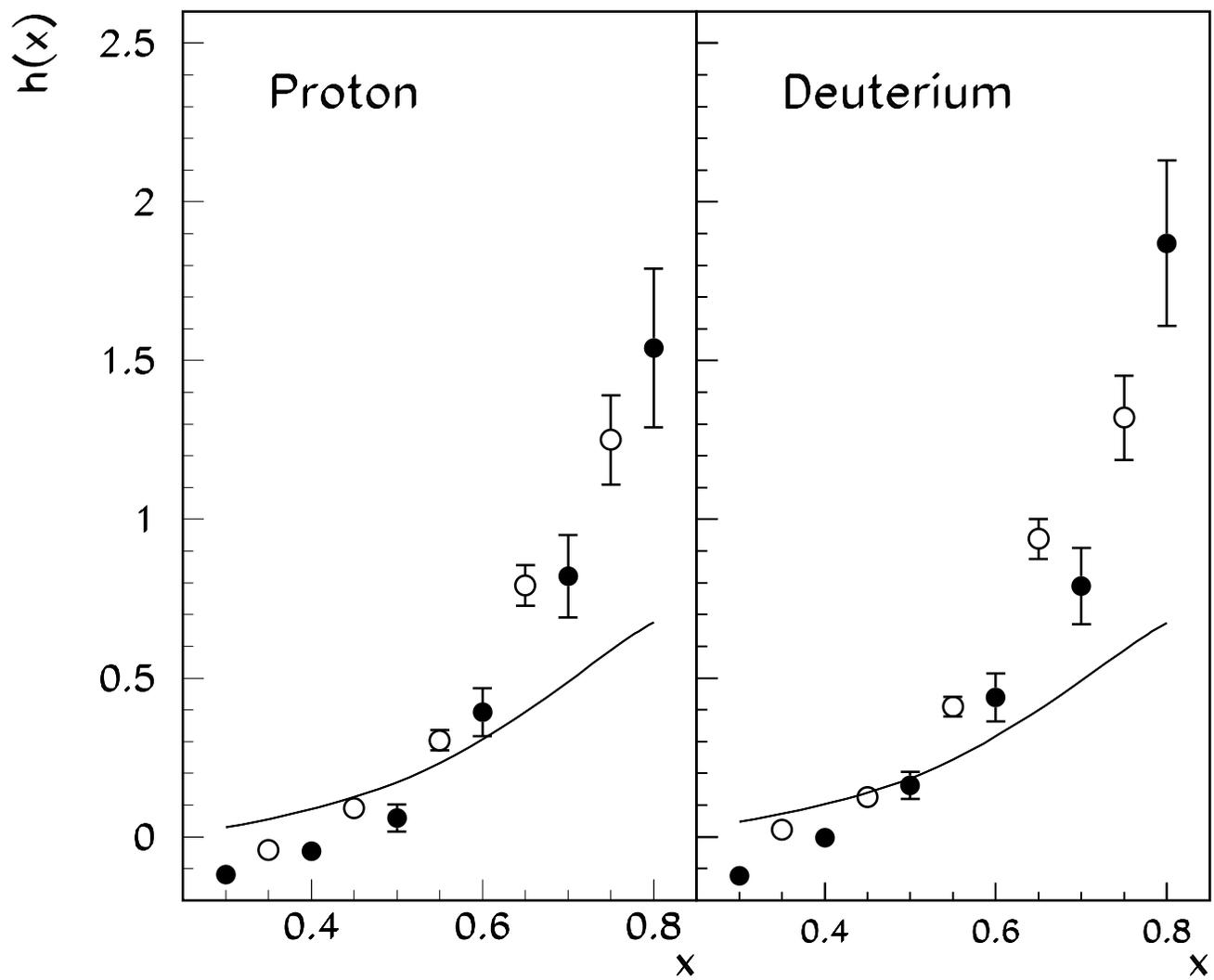
\begin{figure}
\caption{The high-twist contributions obtained in our 
final fit (full circles)
and the results of the analysis [2] (open circles).
The full curves represent the calculations on the IRR model.}
\end{figure}

\newpage
\begin{table}
\caption{The number of data points (NDP) and the number of independent
systematic errors (NSE)
for the analysed data sets.}
\begin{tabular}{cccc} 
Experiment&NDP(proton)&NDP(deuterium)&NSE\\  
BCDMS  &223&162 &9        \\  
E-49A  &47&47   &5   \\  
E-49B  &109&102 &5     \\  
E-61   &6&6     &5 \\  
E-87   &90&90   &5   \\  
E-89A   &66&59  &5    \\  
E-89B  &70&59   &5   \\  
E-139   &--&16  &5    \\  
E-140  &--&31   &4   \\  
TOTAL  &611&572 &45     \\  
\end{tabular}
\end{table}

\begin{table}
\caption{The results of the fits with the various approaches 
to the treatment of the BCDMS systematic errors. 
The parameters $\xi$ and $\lambda$ describe the renormalization and 
shift of the BCDMS data, $h_{3,4,5,6,7,8}$ are the fitted values of 
the HT contribution at $x=0.3,0.4,0.5,0.6,0.7,0.8$.
For the description of the columns see the text.}
\tiny
\begin{tabular}{cccccccc}   
                 &   1               &       2          &   3             &   4              &  5               &   6             &    7             \\ 
$A_p$            & $0.612\pm0.028$   & $0.579\pm0.028$   & $0.581\pm0.028$   & $0.523\pm0.024$   & $0.531\pm0.024$   & $0.531\pm0.024$  & $0.519\pm0.022$    \\ 
$a_p$            & $0.642\pm0.028$   & $0.689\pm0.032$   & $0.685\pm0.032$   & $0.748\pm0.033$   & $0.736\pm0.032$   & $0.734\pm0.032$  & $0.748\pm0.030$    \\ 
$b_p$            & $3.588\pm0.029$   & $3.675\pm0.038$   & $3.670\pm0.038$   & $3.702\pm0.038$   & $3.686\pm0.037$   & $3.670\pm0.037$  & $3.667\pm0.035$    \\ 
$A_n$            & $4.0\pm3.2$       & $4.0\pm3.5$       & $3.7\pm3.0$       & $4.7\pm4.8$       & $3.4\pm2.6$       & $4.2\pm3.8$      & $4.7\pm4.4$        \\ 
$a_n$            & $0.14\pm0.11$     & $0.14\pm0.12$     & $0.15\pm0.12$     & $0.12\pm0.12$     & $0.16\pm0.12$     & $0.13\pm0.12$    & $0.12\pm0.11$      \\ 
$b_n$            & $3.52\pm0.12$     & $3.52\pm0.14$     & $3.54\pm0.14$     & $3.48\pm0.14$     & $3.52\pm0.14$     & $3.48\pm0.14$    & $3.51\pm0.12$      \\ 
$\alpha_s(M_Z)$  & $0.1141\pm0.0007$ & $0.1089\pm0.0016$ & $0.1093\pm0.0016$ & $0.1119\pm0.0015$ & $0.1140\pm0.0017$ & $0.1173\pm0.0018$& $0.1188\pm0.0018$  \\ 
$\lambda_P$      & 1.4               & $0.95\pm0.13$     & $0.97\pm0.13$     & --                & --                & --               & --                 \\ 
$\lambda_D$      & 1.2               & $0.89\pm0.15$     & $0.90\pm0.15$     & --                & --                & --               & --                 \\ 
$\xi_P$          & 0.99              & $1.0138\pm0.0059$ & --                & --                & --                & --               & --                 \\ 
$\xi_D$          & 1.004             & $1.0261\pm0.0063$ & --                & --                & --                & --               & --                 \\ 
$h^P_3$          & $-0.154\pm0.016$  & $-0.136\pm0.017$  & $-0.138\pm0.016$  & $-0.114\pm0.017$  & $-0.125\pm0.018$  & $-0.136\pm0.018$ & $-0.136\pm0.017$   \\ 
$h^P_4$          & $-0.009\pm0.019$  & $0.030\pm0.022$   & $0.026\pm0.022$   & $0.015\pm0.022$   & $-0.010\pm0.024$  & $-0.047\pm0.026$ & $-0.068\pm0.026$   \\ 
$h^P_5$          & $0.175\pm0.029$   & $0.257\pm0.038$   & $0.250\pm0.037$   & $0.191\pm0.038$   & $0.149\pm0.041$   & $0.077\pm0.045$  & $0.029\pm0.046$    \\ 
$h^P_6$          & $0.623\pm0.054$   & $0.803\pm0.072$   & $0.788\pm0.070$   & $0.643\pm0.071$   & $0.572\pm0.077$   & $0.440\pm0.083$  & $0.338\pm0.084$    \\ 
$h^P_7$          & $1.106\pm0.089$   & $1.49\pm0.13$     & $1.46\pm0.13$     & $1.23\pm0.13$     & $1.11\pm0.13$     & $0.90\pm0.14$    & $0.73\pm0.14$      \\ 
$h^P_8$          & $1.83\pm0.25$     & $2.56\pm0.31$     & $2.51\pm0.31$     & $2.20\pm0.30$     & $1.99\pm0.30$     & $1.66\pm0.31$    & $1.41\pm0.30$      \\ 
$h^D_3$          & $-0.130\pm0.018$  & $-0.102\pm0.019$  & $-0.103\pm0.019$  & $-0.094\pm0.019$  & $-0.102\pm0.020$  & $-0.123\pm0.021$ & $-0.129\pm0.021$   \\ 
$h^D_4$          & $0.048\pm0.017$   & $0.104\pm0.022$   & $0.099\pm0.022$   & $0.081\pm0.022$   & $0.054\pm0.025$   & $0.010\pm0.028$  & $-0.005\pm0.029$   \\ 
$h^D_5$          & $0.266\pm0.027$   & $0.367\pm0.038$   & $0.358\pm0.037$   & $0.299\pm0.038$   & $0.248\pm0.042$   & $0.172\pm0.047$  & $0.146\pm0.049$    \\ 
$h^D_6$          & $0.657\pm0.050$   & $0.844\pm0.069$   & $0.829\pm0.068$   & $0.696\pm0.068$   & $0.611\pm0.075$   & $0.480\pm0.082$  & $0.445\pm0.085$    \\ 
$h^D_7$          & $1.050\pm0.075$   & $1.38\pm0.11$     & $1.36\pm0.11$     & $1.15\pm0.11$     & $1.03\pm0.12$     & $0.82\pm0.13$    & $0.77\pm0.13$      \\ 
$h^D_8$          & $2.28\pm0.25$     & $2.96\pm0.31$     & $2.92\pm0.31$     & $2.52\pm0.30$     & $2.34\pm0.30$     & $1.98\pm0.31$    & $1.94\pm0.31$      \\ 
$\chi^2$         & $1090.5$          & $1067.5$          & $1068.3$          & $963.7$           & $964.3$           & $973.3$          & $971.5$            \\ 
\end{tabular}
\normalsize
\end{table}

\begin{table}
\caption{The results of the fits with various approaches 
to the treatment of the SLAC systematic errors. 
The parameters $\xi$ describe the renormalization
of the SLAC data, $h_{3,4,5,6,7,8}$ are the fitted values of 
the HT contribution at $x=0.3,0.4,0.5,0.6,0.7,0.8$.
For the description of the columns see the text.}
\begin{tabular}{cccc}      
                 &   1               &       2          &       3         \\ 
$A_p$            & $0.527\pm0.022$   & $0.546\pm0.025$   & $0.516\pm0.022$   \\ 
$a_p$            & $0.738\pm0.030$   & $0.723\pm0.030$   & $0.765\pm0.028$   \\ 
$b_p$            & $3.656\pm0.035$   & $3.642\pm0.034$   & $3.692\pm0.032$   \\ 
$A_n$            & $3.8\pm2.9$       & $4.9\pm4.4$       & $4.8\pm4.1$       \\ 
$a_n$            & $0.15\pm0.11$     & $0.12\pm0.10$     & $0.118\pm0.097$   \\ 
$b_n$            & $3.54\pm0.12$     & $3.51\pm0.12$     & $3.51\pm0.11$     \\ 
$\alpha_s(M_Z)$  & $0.1188\pm0.0018$ & $0.1183\pm0.0017$ & $0.1180\pm0.0017$ \\ 
$h^P_3$          & $-0.140\pm0.017$  & $-0.136\pm0.018$  & $-0.120\pm0.017$  \\ 
$h^P_4$          & $-0.069\pm0.026$  & $-0.052\pm0.027$  & $-0.046\pm0.025$  \\ 
$h^P_5$          & $0.031\pm0.046$   & $0.059\pm0.045$   & $0.059\pm0.043$   \\ 
$h^P_6$          & $0.341\pm0.083$   & $0.400\pm0.081$   & $0.392\pm0.076$   \\ 
$h^P_7$          & $0.72\pm0.14$     & $0.79\pm0.13$     & $0.82\pm0.13$     \\ 
$h^P_8$          & $1.38\pm0.30$     & $1.44\pm0.28$     & $1.54\pm0.25$     \\ 
$h^D_3$          & $-0.128\pm0.021$  & $-0.134\pm0.019$  & $-0.123\pm0.018$  \\ 
$h^D_4$          & $-0.005\pm0.029$  & $-0.007\pm0.027$  & $-0.003\pm0.026$  \\ 
$h^D_5$          & $0.145\pm0.049$   & $0.159\pm0.045$   & $0.162\pm0.043$   \\ 
$h^D_6$          & $0.442\pm0.084$   & $0.446\pm0.080$   & $0.439\pm0.076$   \\ 
$h^D_7$          & $0.79\pm0.13$     & $0.77\pm0.12$     & $0.79\pm0.12$     \\ 
$h^D_8$          & $1.93\pm0.31$     & $1.84\pm0.29$     & $1.87\pm0.26$     \\ 
$\xi_{P,49A}$    & --                & $1.016\pm0.017$   & $1.016\pm0.018$   \\ 
$\xi_{D,49A}$    & --                & $1.007\pm0.016$   & $1.006\pm0.017$   \\ 
$\xi_{P,49B}$    & --                & $1.021\pm0.017$   & $1.028\pm0.018$   \\ 
$\xi_{D,49B}$    & --                & $1.006\pm0.016$   & $1.012\pm0.017$   \\ 
$\xi_{P,61}$     & --                & $1.019\pm0.020$   & $1.021\pm0.021$   \\ 
$\xi_{D,61}$     & --                & $1.004\pm0.018$   & $1.004\pm0.019$   \\ 
$\xi_{P,87}$     & --                & $1.018\pm0.017$   & $1.025\pm0.017$   \\ 
$\xi_{D,87}$     & --                & $1.006\pm0.016$   & $1.012\pm0.017$   \\ 
$\xi_{P,89A}$    & --                & $1.023\pm0.018$   & $1.028\pm0.021$   \\ 
$\xi_{D,89A}$    & --                & $1.001\pm0.017$   & $1.004\pm0.021$   \\ 
$\xi_{P,89B}$    & --                & $1.022\pm0.017$   & $1.022\pm0.017$   \\ 
$\xi_{D,89B}$    & --                & $1.007\pm0.016$   & $1.007\pm0.017$   \\ 
$\xi_{D,139}$    & --                & $1.012\pm0.016$   & $1.009\pm0.017$   \\ 
$\chi^2$         & $971.8$           & $1040.8$          & $1178.9$          \\ 
\end{tabular}
\end{table}

\begin{table}
\caption{The correlation matrix for the parameters from the final fit.}
\tiny
\begin{tabular}{cccccccccccccccccccc}   
    &$a_p$ & $b_p$ & $a_n$ & $b_n$ & $\alpha_s(M_Z)$ & $A_p$ & $A_n$ &$h^P_3$&$h^P_4$&$h^P_5$&$h^P_6$&$h^P_7$&$h^P_8$&
$h^D_3$&$h^D_4$&$h^D_5$&$h^D_6$&$h^D_7$&$h^D_8$ \\ 
$a_p$  & 1.00&0.93&-0.50&-0.45&-0.07&-0.92&0.50&0.61&0.11&-0.10&-0.09&0.12&0.34&0.01&0.08&0.05&0.05&0.08&0.11   \\ 
$b_p$  &0.93&1.00&-0.46&-0.44&-0.27&-0.82&0.46&0.55&0.22&0.09&0.14&0.37&0.53&0.11&0.25&0.25&0.24&0.28&0.25      \\ 
$a_n$  &-0.50&-0.46&1.00&0.96&-0.04&0.46&-0.99&-0.28&0.01&0.12&0.12&-0.01&-0.14&0.56&0.04&-0.07&-0.04&0.11&0.28 \\ 
$a_n$  &-0.45&-0.44&0.96&1.00&-0.09&0.40&-0.96&-0.19&0.07&0.16&0.13&0.00&-0.14&0.51&0.06&-0.02&0.04&0.12&0.36   \\ 
$\alpha_s(M_Z)$ &-0.07&-0.27&-0.04&-0.09&1.00&0.04&0.04&-0.33&-0.79&-0.88&-0.89&-0.87&-0.67&-0.55&-0.90&-0.95&-0.92&-0.92&-0.67 \\ 
$A_p$  &-0.92&-0.82&0.46&0.40&0.04&1.00&-0.46&-0.58&-0.11&0.11&0.11&-0.07&-0.28&-0.01&-0.06&-0.04&-0.03&-0.05&-0.08 \\ 
$A_n$  &0.50&0.46&-0.99&-0.96&0.04&-0.46&1.00&0.28&-0.00&-0.12&-0.12&0.00&0.14&-0.56&-0.04&0.07&0.04&-0.11&-0.27 \\ 
$h^P_3$&0.61&0.55&-0.28&-0.19&-0.33&-0.58&0.28&1.00&0.40&0.34&0.25&0.32&0.35&0.20&0.35&0.34&0.31&0.32&0.25   \\ 
$h^P_4$&0.11&0.22&0.01&0.07&-0.79&-0.11&-0.00&0.40&1.00&0.76&0.78&0.71&0.54&0.45&0.75&0.77&0.74&0.73&0.52 \\ 
$h^P_5$&-0.10&0.09&0.12&0.16&-0.88&0.11&-0.12&0.34&0.76&1.00&0.84&0.81&0.56&0.50&0.81&0.85&0.82&0.81&0.58  \\ 
$h^P_6$&-0.09&0.14&0.12&0.13&-0.89&0.11&-0.12&0.25&0.78&0.84&1.00&0.82&0.65&0.50&0.81&0.86&0.84&0.83&0.60  \\ 
$h^P_7$&0.12&0.37&-0.01&0.00&-0.87&-0.07&0.00&0.32&0.71&0.81&0.83&1.00&0.64&0.48&0.79&0.83&0.82&0.82&0.61  \\ 
$h^P_8$&0.34&0.53&-0.14&-0.14&-0.67&-0.28&0.14&0.35&0.54&0.56&0.65&0.64&1.00&0.35&0.61&0.64&0.63&0.64&0.50  \\ 
$h^D_3$&0.01&0.11&0.56&0.51&-0.55&-0.01&-0.56&0.20&0.45&0.50&0.50&0.48&0.35&1.00&0.51&0.50&0.44&0.52&0.50  \\ 
$h^D_4$&0.08&0.25&0.04&0.06&-0.90&-0.06&-0.04&0.35&0.75&0.81&0.81&0.79&0.61&0.51&1.00&0.90&0.86&0.83&0.59  \\ 
$h^D_5$&0.05&0.25&-0.07&-0.02&-0.95&-0.04&0.07&0.34&0.77&0.85&0.86&0.83&0.64&0.50&0.90&1.00&0.92&0.89&0.60  \\ 
$h^D_6$&0.05&0.24&-0.04&0.04&-0.92&-0.03&0.04&0.31&0.74&0.82&0.84&0.82&0.63&0.44&0.86&0.92&1.00&0.89&0.66  \\ 
$h^D_7$&0.08&0.28&0.11&0.20&-0.92&-0.05&-0.11&0.32&0.73&0.81&0.83&0.82&0.64&0.52&0.83&0.89&0.89&1.00&0.68  \\ 
$h^D_8$&0.11&0.25&0.28&0.36&-0.67&-0.08&-0.27&0.25&0.52&0.58&0.60&0.61&0.50&0.50&0.59&0.60&0.66&0.68&1.00 \\ 
\end{tabular}
\normalsize
\end{table}

\end{document}